\documentclass[12pt]{article}
\begin{document}

\title{{\bf Quantum Uncertainties in the Schmidt Basis Given by
Decoherence}}

\author{
Don N. Page
\thanks{Internet address:
profdonpage@gmail.com}
\\
Department of Physics\\
4-183 CCIS\\
University of Alberta\\
Edmonton, Alberta T6G 2E1\\
Canada
}

\date{2011 August 9}

\maketitle
\large
\begin{abstract}
\baselineskip 25 pt

A common misconception is that decoherence gives the eigenstates that we
observe to be fairly definite about a subsystem (e.g., approximate
eigenstates of position) as the elements of the Schmidt basis in which
the density matrix of the subsystem is diagonal.  Here I show that in
simple examples of linear systems with gaussian states, the Schmidt
basis states have as much mean uncertainty about position as the full
density matrix with its combination of different possibilities.

\end{abstract}

\normalsize

\baselineskip 23.5 pt

\newpage

A question puzzling some of us scientists is why we our observations are
fairly definite, and why what appears to us visually to be be fairly
definite is usually something like approximate positions of objects that
we see.  One might say that what we see is presumably determined by the
firings of neurons in the retinas at the backs of our eyes, so that a
particular pattern of firings gives a visual impression of the locations
of objects that we see.  Then the question is why we are visually aware
of a fairly definite pattern of retinal neuron firings.  One might go on
to say that this is because a fairly definite pattern of these retinal
neuron firings induces a fairly definite pattern of neuron firings or
some other property in some more central part of the brain where the
visual awareness may be postulated to occur, but this just pushes the
question back to why we are aware of those brain properties, rather than
of superpositions of them.  If we assume that for particular brain
properties, there are corresponding visual awarenesses of objects that
appear to have fairly definite positions, rather than of combinations of
different positions, the question is then what is the preferred basis of
states for the subsystem of these brain properties, such that each basis
state leads mainly to a single definite visual awareness.

The mystery arises because in quantum theory, unitary evolution would
almost always lead to a state of the brain subsystem that is a mixed
state of the particular brain properties that each lead to fairly
definite visual awarenesses.  If the brain state in quantum theory is a
mixed state of many brain properties, what picks out the particular
brain states that each lead mainly to a fairly definite observation that
one is aware of having?  Or, if we assume that the process of vision
maps the relative positional configuration of an observed object to a
corresponding brain property, what picks out these particular states of
the object (rather than superpositions of them) that each lead to a
fairly definite visual awareness of the object?  Observationally, these
seem to be approximate position eigenstates of the object, but why is
that true?

Traditionally it was postulated that the quantum state (of a closed
system) not only has the unitary evolution of the Schr\"{o}dinger
equation, but that at certain times the unitary evolution is broken by
the so-called ``collapse of the wavefunction'' to return it to a
macroscopically definite quantum state \cite{vN}, such as an approximate
position eigenstate of observed objects.  This collapse was supposed to
occur during measurements, but usually it was left rather vague what
precisely constituted a measurement and exactly when the collapse of the
wavefunction would occur.

More recently it has become widely recognized that quantum subsystems of
the universe rapidly become entangled with their environments through
generic interaction processes called decoherence, so that the subsystems
are not in pure states but mixed states, described by density matrices
or density operators that are not the unit-rank projection operators
that are the density operators of pure quantum states
\cite{Zeh70,Zeh73,Zurek81,Zurek82,JZ,Zurek86,Zurek91,Albrecht92,PHZ,
Albrecht93,ZHP,Zurek93,TS,PZ1,ZP,GKZ,Gallis,AZ,APZ,Zurek98a,HSZ,
Zurek98b,PZ2,Zeh99,Zeh00,BGJKS,Zurek03,JZKGKS,Schl1,Zeh05,Schl2,KZ,
Zeh10}.   It is sometimes assumed that it is the eigenstates of the
density operator of a subsystem produced by decoherence (see, e.g.,
\cite{Zeh73,Albrecht92,Albrecht93}, who use these eigenstates) that are
the particular states of a subsystem that each lead to fairly definite
visual awarenesses.  For example, I got that impression from some
recent statements of Raphael Bousso and Leonard Susskind \cite{BS}, in
a paper that has several other interesting ideas whose truth or
falsehood seems to be rather independent of how I interpreted their
statements about decoherence.  They wrote, ``Decoherence explains why
observers do not experience superpositions of macroscopically distinct
quantum states, such as a superposition of an alive and a dead cat.
\ldots\  Decoherence explains the `collapse of the wave function' of the
Copenhagen interpretation as the non-unitary evolution from a pure to a
mixed state, resulting from ignorance about an entangled subsystem $E$. 
It also explains the very special quantum states of macroscopic objects
we experience, as the elements of the basis in which the density matrix
$\rho_{SA}$ is diagonal."

Here I wish to correct the misconception that seemed to me to be
implicit in the last sentence above, that the eigenstates of the
subsystem density matrix (the Schmidt basis for it) each have the
variables that are observed to be macroscopically definite (e.g.,
positions) in macroscopically definite states.  It is not generically
true that the Schmidt basis, in which the subsystem density matrix is
diagonal, would have basis states (the eigenstates of the subsystem
density matrix) in each of which there is a single macroscopic state,
such as the approximate position eigenstates that we appear to observe
in each individual visual observation.  It may be true that with
interactions that are local in space, the density matrix in a basis that
each has an appropriate single macroscopic state (e.g., an appropriate
superposition of quantum microstates that each have the same unique
macroscopic values) is often approximately diagonal, but the basis in
which the density matrix really is precisely diagonal is, as I shall
show for a wide class of simple examples, far from each having definite
macroscopic states.  In particular, I shall show that for many simple
examples the mean uncertainty of the position variables in each of the
Schmidt basis states is just as great as the full uncertainty that they
have in the complete quantum density matrix of the subsystem.

To quantify how much mean uncertainty there is in a certain operator
(e.g., for a macroscopic variable such as center-of-mass position) in a
certain basis (e.g., the Schmidt basis), I shall define a dimensionless
{\it Mean Observable-Basis Uncertainty} (MOBU) $U_{OB}$ as the ratio of
the mean variance of the observable $O$ (a hermitian operator) in a
basis $B ={|i>}$ ($i=1\ldots m$) of $m$ pure states for a quantum
subsystem of Hilbert-space dimension $n$ and with a mixed state given by
the density operator $\rho$, to the full variance of the observable in
the same mixed state.  If $B$ is the Schmidt basis, one has that $m=n$
and that the ${|i\rangle}$ are the orthonormal eigenvectors of the
density operator, which can be written as $\rho = \sum_{j=1}^n
p_j|j\rangle\langle j|$ with nonnegative eigenvalues $p_j$ that sum to
unity (and which are often interpreted to be the probabilities that the
quantum subsystem is in each of the $n$ pure states $|j\rangle$ that are
the orthonormal eigenvectors of the density operator).  However, I shall
give a general definition of the MOBU $U_{OB}$ for an arbitrary basis
$B$, without even assuming that the $m$ ${|i\rangle}$ are orthonormal.

The full variance of the observable $O$ in the (normalized) density
matrix of the quantum subsystem is $(\Delta O)^2 \equiv
\langle(O-\langle O\rangle)^2\rangle = \langle O^2\rangle - \langle
O\rangle^2 \equiv \mathrm{tr}(\rho O^2) - [ \mathrm{tr}(\rho O)]^2$. 
The variance in the pure state $|i\rangle$ (not necessarily assumed to
be normalized) is $(\Delta O_i)^2 \equiv \langle i|O^2|i\rangle/\langle
i|i\rangle - (\langle i|O|i\rangle/\langle i|i\rangle)^2$.  Define the
probabilities $P_i$ for the basis states ${|i\rangle}$, given the
subsystem density operator $\rho$, to be $P_i \equiv r_i/N$ with $r_i
\equiv \langle i|\rho|i\rangle/\langle i|i\rangle$ being the relative
probability for the basis state $|i\rangle$ and with $N \equiv
\sum_{i=1}^m r_i$ being the normalization factor, which will be unity if
$m=n$ and if the ${|i\rangle}$ are orthogonal. For the Schmidt basis
${|j\rangle}$ of $m=n$ orthonormal eigenvectors of the density operator
$\rho$, $P_j = r_j = \langle j|\rho|j\rangle = p_j$, but for a generic
basis I shall reserve $p_j$ for the $n$ eigenvalues of $\rho$ and use
$P_i$ for the probabilities of the $m$ basis states ${|i\rangle}$.  Then
the subsystem has a mean variance of the observable $O$ in the basis $B
= {|i\rangle}$ that can be defined to be $(\Delta O_B)^2 \equiv
\sum_{i=1}^m P_i(\Delta O_i)^2$.  Finally, define the dimensionless {\it
Mean Observable-Basis Uncertainty} (MOBU) as $U_{OB} \equiv (\Delta
O_B)^2/(\Delta O)^2$, the ratio of the mean variance of the observable
$O$ in the basis $B$ to the full variance of $O$.

Given an observable $O$ that represents what is believed to be observed
to have definite values (e.g., macroscopic positions), a goal would be
to find a basis of states $B ={|i>}$ that gives a small value of the
Mean Observable-Basis Uncertainty $U_{OB}$.  Of course, one can just
choose an orthonormal basis of eigenvectors of $O$, and then in each
pure eigenstate $|i\rangle$ of $O$, $(\Delta O_i)^2 = 0$, so the mean
variance of the observable in this basis is $(\Delta O_B)^2 = 0$, giving
$U_{OB} = 0$ if the full variance $(\Delta O)^2$ of $O$ in the density
operator of the quantum subsystem is positive.  However, the question
here is whether one gets a small $U_{OB}$ from the Schmidt basis of
eigenvectors of the subsystem density operator $\rho$.

One can easily see that if there is no restriction on the basis (even if
it is required to be an orthonormal basis of the same dimension $n$ as
the Hilbert space of the subsystem under consideration), then the MOBU
can be any positive number (including infinity, if the observable has no
full variance in the density operator of the subsystem).  However, it is
restricted to be no greater than unity for the Schmidt basis
${|i\rangle} = {|j\rangle}$, as I shall now show.

Assuming that $B ={|i>}$ is the Schmidt basis of $m=n$ orthonormal
eigenstates of the density operator, which can then be written as $\rho
= \sum_{i=1}^m P_i|i\rangle\langle i|$ with nonnegative eigenvalues $P_i
= p_i$ that are the same as what was defined above to be the
probabilities for these particular basis states, define the mean value
of the observable $O$ in the pure state $|i\rangle$ to be $O_i \equiv
\langle i|O|i\rangle$ and the mean value in the full mixed state to be
$O_m \equiv \langle O\rangle \equiv \mathrm{tr}(\rho O) = \sum_{i=1}^n
p_i \langle i|O|i\rangle = \sum_{i=1}^m P_i O_i$.  Interpret $O_i$ and
$O_m$ to mean these expectation values multiplied by the identity
operator when they are used inside quantum inner products.  Then the
variance of the observable $O$ in the normalized pure state $|i\rangle$
is $(\Delta O_i)^2 = \langle i|(O-O_i)^2|i\rangle$, the mean variance of
$O$ in the Schmidt basis $B$ is $(\Delta O_B)^2 = \sum_{i=1}^m P_i
\langle i|(O-O_i)^2|i\rangle$, and the full variance of $O$ in the mixed
state of the subsystem is $(\Delta O)^2 = \sum_{i=1}^m P_i \langle
i|(O-O_m)^2|i\rangle$, which is greater than the mean variance by the
nonnegative excess $E = (\Delta O)^2 - (\Delta O_B)^2 =  \sum_{i=1}^m
P_i \langle i|[(O-O_m)^2 - (O-O_i)^2]|i\rangle =  \sum_{i=1}^m P_i
\langle i|(O_i-O_m)(2O-O_i-O_m)|i\rangle = \sum_{i=1}^m P_i(O_i-O_m)^2
\geq 0$.  Therefore, the Mean Observable-Basis Uncertainty is $U_{OB}
\equiv (\Delta O_B)^2/(\Delta O)^2 = 1 - E/(\Delta O_B)^2 \leq 1$ for
the Schmidt basis.

Thus the Schmidt basis never gives a mean variance of an observable $O$
in its basis states that is larger than the full variance of $O$, unlike
what is possible with other bases.  However, it can give a mean variance
as large as the full variance, and hence a MOBU value of $U_{OB} = 1$,
if the mean value of $O$ in each Schmidt basis state is the same, $O_i =
O_m$ for all $i$ for which $P_i > 0$.  Next I shall show that this is
what indeed occurs for the position observable in simple models of
linear coupling of harmonic oscillators in which the subsystem density
matrix has a gaussian form as the exponential of a negative quadratic
expression in the positions and/or momenta.

There has been an extensive study of quantum models in which a free
particle or harmonic oscillator (which I shall take as the quantum
subsystem of interest) interacts linearly with a collection of other
harmonic oscillators (which I shall call the environment; in some cases
it may be taken to be a heat bath)
\cite{Nak,Haake,Haken,Dekker,CL,HR,UZ}.  For simplicity, one often takes
the initial density matrix for the whole coupled system to have a
gaussian form, such as a product (no initial entanglement) of a pure or
mixed gaussian state for the subsystem of interest and another pure or
mixed state for the environment (such as a thermal state).  The linear
coupling leads to entanglement between the subsystem and the
environment, but for free particles or harmonic oscillators with linear
couplings between the positions and/or momenta, the full quantum state
retains its gaussian form (proportional to the exponential of a negative
quadratic expression in all the positions and/or momenta).  It is then
easy to see that tracing over the environment gives a gaussian density
matrix for the subsystem of interest \cite{Feynman}.

By performing a canonical transformation of the position and momentum
operators (which generically includes not only rotations in the phase
space but also shifting the expectation values for the transformed
positions and momenta to zero and rescaling the positions and momenta by
a squeeze), one can write the gaussian density matrix as a product of
thermal gaussian states for each degree of freedom for the transformed
system, with thermality defined with respect to a `Hamiltonian' that for
each transformed degree of freedom is half the sum of the squares of the
transformed position and momentum variables \cite{UZ}.  The Schmidt
basis of eigenstates of this subsystem density matrix are then the
products of the energy eigenstates of this `Hamiltonian' for each degree
of freedom.  Each of these eigenstates has zero expectation values for
each of the transformed position and momentum variables.  Hence, it has
the same values (though generically not zero) for the expectation values
of the original position and momentum variables.  Therefore, for any
observable $O$ that is a linear combination of position and momentum
variables, the Schmidt basis for such a gaussian state of the subsystem
gives the same value for the expectation value $O_i \equiv \langle
i|O|i\rangle$ for each Schmidt basis state, $O_i = O_m = \langle
O\rangle = \mathrm{tr}(\rho O)$.  By the result above, this leads to the
excess of the full variance of $O$ over the mean variance of $O$ in the
Schmidt basis being $E=0$, so the dimensionless Mean Observable-Basis
Uncertainty is $U_{OB} = 1$.

Thus, on average, each eigenstate of the density operator for the
quantum subsystem gives no less uncertainty for a position (or momentum,
or linear combination of position and momentum) observable than the full
density matrix does.  That is, each element of the Schmidt basis given
by decoherence leads to a mean uncertainty in position just as great as
the uncertainty given by the entire density matrix of the subsystem. 
For these common examples of linearly coupled harmonic oscillators in
gaussian states, decoherence does not lead to eigenstates of the
subsystem density operator that are sharp in position.

It would be interesting to analyze quantum systems that are not in
gaussian states, either from imposing nongaussian initial states or from
having nonlinear couplings, to see how the Mean Observable-Basis
Uncertainty behaves.  For example, if one starts with the subsystem
having a superposition of two widely separated gaussian states but
continues to restrict to linear couplings for simplicity, does the MOBU
evolve to become very small, or do the eigenstates of the density matrix
for the subsystem each continue to have significant contributions from
the two widely separated locations to keep the MOBU of the order of
unity?  Such considerations will be left for future research.

It would also be interesting to calculate the MOBU for the more
sophisticated `pointer bases' that have been proposed
\cite{Zurek81,Zurek82,Zurek86,Zurek91,PHZ,Zurek98b,PZ2,Zurek03,Schl1,
Eisert,DDZ,Zurek09,RZ}.  Although none of these have been precisely
defined for a generic situation, they are variously described as ``the
eigenvectors of the operator which commutes with the
apparatus-environment interaction Hamiltonian'' and ``contains a
reliable record of the state of the system'' \cite{Zurek81}, the
eigenstates of the observable of the apparatus ``which is most reliably
recorded by the environment'' \cite{Zurek82}, the states ``which become
minimally entangled with the environment in the course of the
evolution'' \cite{PZ2}, ``the pure states least affected by
decoherence'' \cite{Zurek03}, ``states that produce the least entropy,''
``states that are the easiest to find out from the imprint they leave on
the environment,'' ``states that can be deduced from measurements on the
smallest fraction of the environment,'' and as ``states for which it
takes the longest to lose a set fraction their initial purity''
\cite{DDZ}.  It is also admitted that ``There is no {\it a priori}
reason to expect that all of these criteria will lead to identical sets
of preferred states,'' though it is ``reasonable to hope that, in the
macroscopic limit in which classicality is indeed expected, differences
between various sieves should be negligible'' \cite{Zurek03}.  I do
suspect that often the MOBU for the position observable would be rather
small for pointer bases that are suitably defined, but that remains to
be calculated.

Because of the ambiguity of which of the many qualitative criteria to
choose for pointer bases, and of how to make any of them precise, I do
not think these many different ideas about pointer bases are the final
answer to the question of how to explain our observations, though they
do seem to be important steps in the right direction.  To me it appears
simplest to postulate that measures or probabilities for our
observations are given by the expectation values of certain definite
`awareness operators,' but we do not yet know what they are and how the
contents of our observations may correlate with these operators
themselves \cite{SQM,MS,CQ}.

In conclusion, each fairly definite location that we observe visually
for an object does not appear to have the form given by any of the
eigenstates of the subsystem density operator after decoherence, at
least for linearly interacting systems with gaussian density operators. 
So this simple-minded idea from decoherence (already criticized in
\cite{PHZ,Zurek98b,Schl1}) seems somewhat incoherent.

I have benefited from recent discussions on this subject with Andy
Albrecht, Raphael Bousso, Sean Carroll, Brandon Carter, David Deutsch,
Jim Hartle (who particularly gave me many useful comments on an early
draft of this paper), John Leslie, Juan Maldacena, Joe Polchinski,
Martin Rees, Rafael Sorkin, Lenny Susskind, Max Tegmark, Bill Unruh, Bob
Wald, and Wojciech Zurek (who painstakingly provided me with many
explanations of the pointer bases and helpful references on these bases
and on decoherence), some of which occurred at the Peyresq Physics 16
symposium of the Peyresq Foyer d'Humanisme in Peyresq, France, 2011 June
18-24, under the hospitality of Edgard Gunzig and OLAM, Association pour
la Recherche Fondamentale, Bruxelles.  I also appreciated the
hospitality of the Perimeter Institute for Theoretical Physics in
Waterloo, Ontario, Canada, where the calculations were made and this
paper was written up, and where I had many more discussions about it. 
This work was supported in part by the Natural Sciences and Engineering
Research Council of Canada.

\end{document}